\documentclass[twocolumns]{sdl2}

\def\0{\mbox{\tiny $0$}}
\def\1{\mbox{\tiny $1$}}
\def\2{\mbox{\tiny $2$}}
\def\3{\mbox{\tiny $3$}}
\def\4{\mbox{\tiny $4$}}
\def\5{\mbox{\tiny $5$}}
\def\6{\mbox{\tiny $6$}}
\def\7{\mbox{\tiny $7$}}
\def\8{\mbox{\tiny $8$}}
\def\9{\mbox{\tiny $9$}}
\def\m{\mbox{\tiny $-$}}
\def\o{\mbox{\tiny $o$}}

\def\wo{\mathrm{w}_{\0}}
\def\w{\mathrm{w}}
\def\d{\mathrm{d}}

\def\xt{\widetilde{x}}

\def\zt{\widetilde{z}}
\def\xs{x_*}

\def\zs{z_*}

\def\inc{_{_{\mathrm{INC}}}}

\def\U{_{_{\mathrm U}}}
\def\L{_{_{\mathrm L}}}
\def\sym{_{_{\mathrm{sym}}}}
\def\cri{_{_{\mathrm{cri}}}}

\def\figureone{
\WideFigureSideCaption{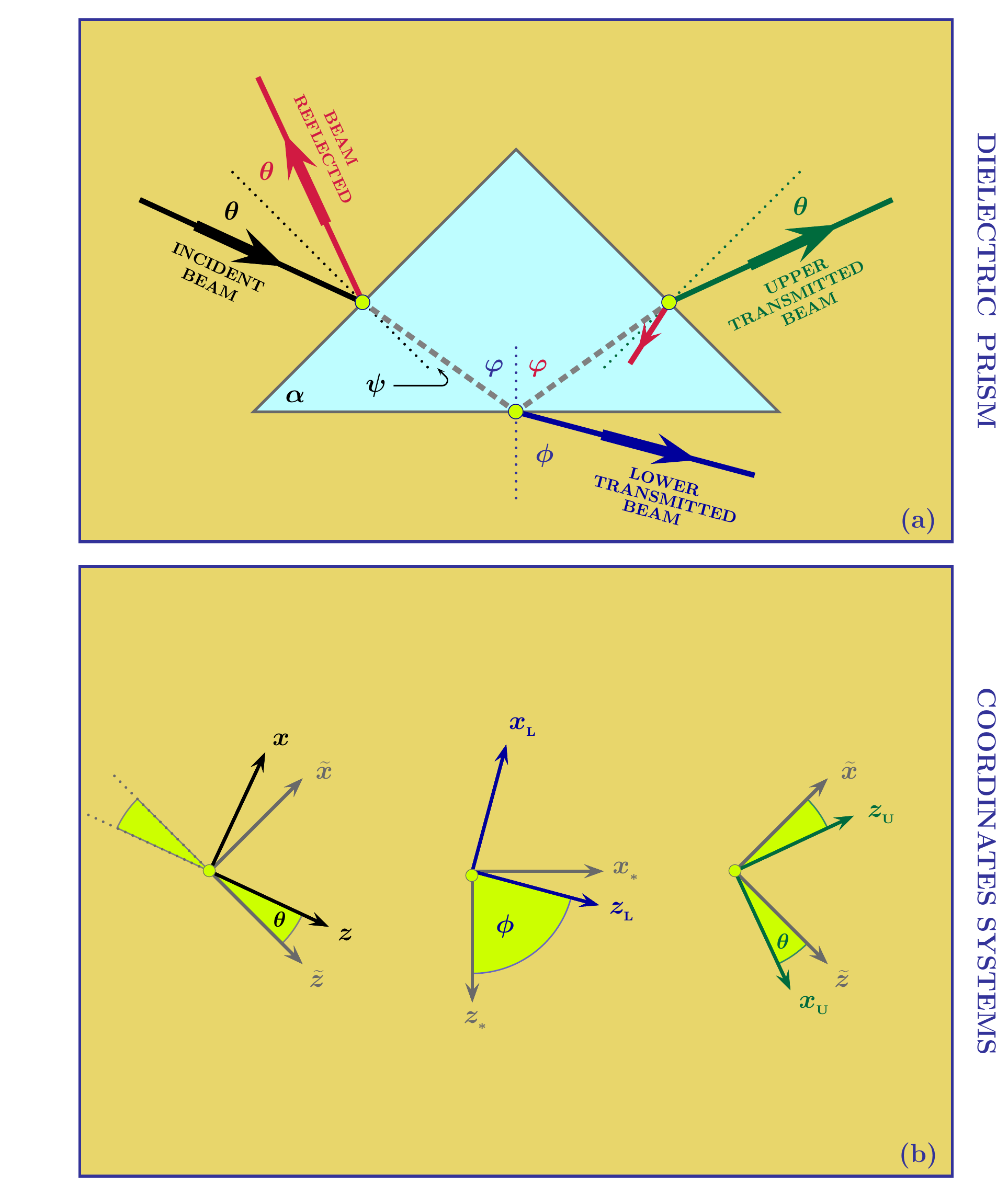}{Geometrical layout of the dielectric used in the study of the lower transmitted beam. In (a) a dielectric prism with an angle $\alpha=\pi/4$, In (b) the system of coordinates of the incident, lower transmitted, and upper transmitted beams. The $\zt$ and $\zs$ axes represent the normal to the left (air/dielectric) and to the lower (dielectric/air) interfaces. The $z$, $z\L$, and  $z\U$ axes indicate the propagation direction of the incident, lower transmitted, and upper transmitted beams. The angles $\psi$ and 
$\phi$ contain an implicit dependence on the incidence angles  at the left ($\theta$) and  lower ($\varphi$) interface, by the Snell law, i.e. $\sin\theta=n\,\sin\psi$ and $n\,\sin\varphi=\sin\phi$.  Finally, the angle $\varphi$ contains the information of the geometrical properties of the dielectric, $\varphi=\psi+\alpha$.         \label{fig1}}
}

\def\figuretwo{
\WideFigureSideCaption{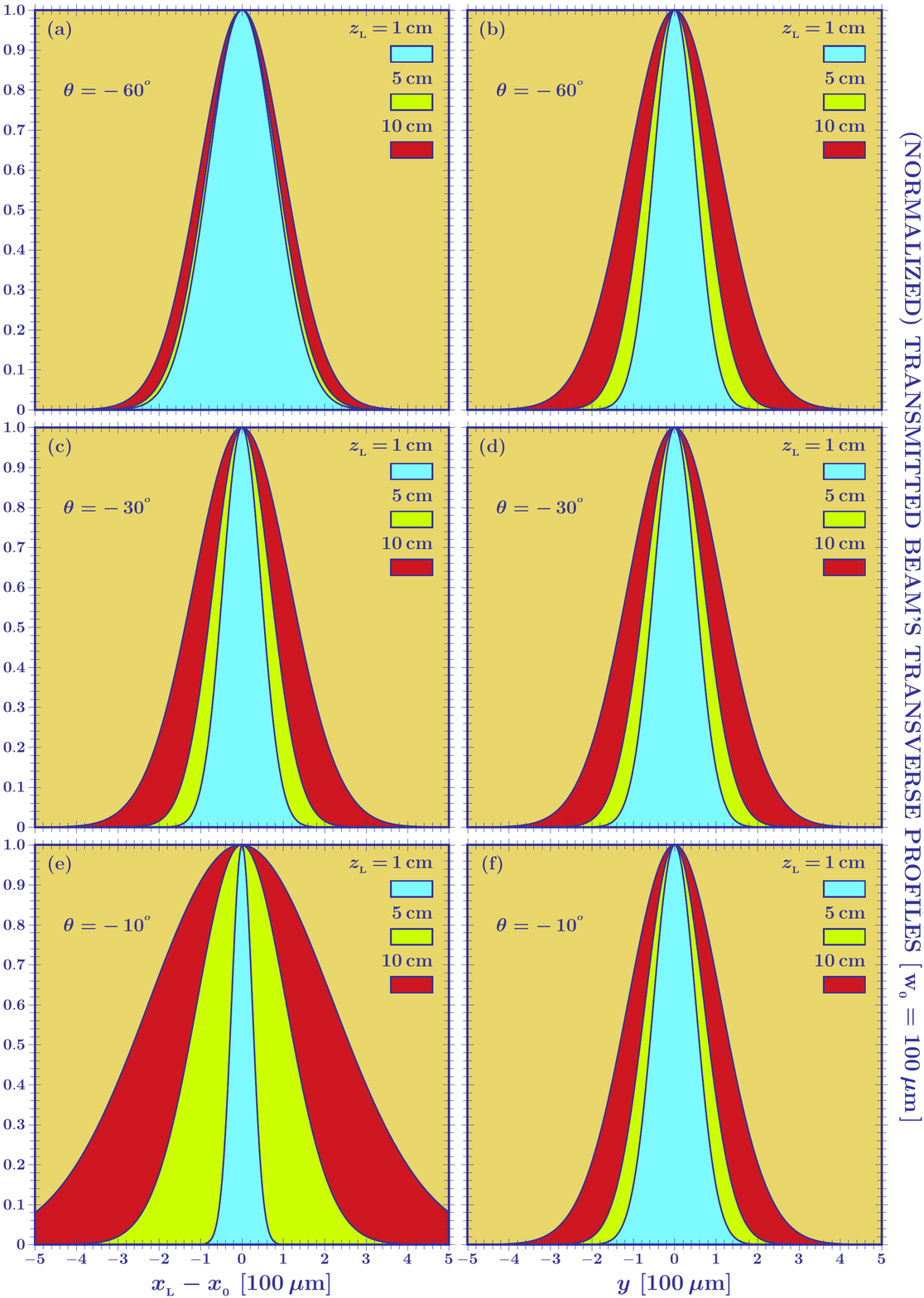}{Transverse profiles of the lower transmitted beam for an incident Gaussian laser with wavelength of $633$ nm and a beam waist of $100\,\mu$m, are plotted  for different incidence angles ($-\,10^{^{o}}$,  $-\,30^{^{o}}$,  and  $-\,60^{^{o}}$) and different axial positions ($1$, $5$, and $10$ cm). 
The $y$ profiles do not change with the incidence angle. The $x\L$ profile show an amplification of the waist 
for $\theta=-\,60^{^{o}}$ and a reduction for $\theta=-\,30^{^{o}}$.   \label{fig2}}}

\def\figurethree{
\WideFigureSideCaption{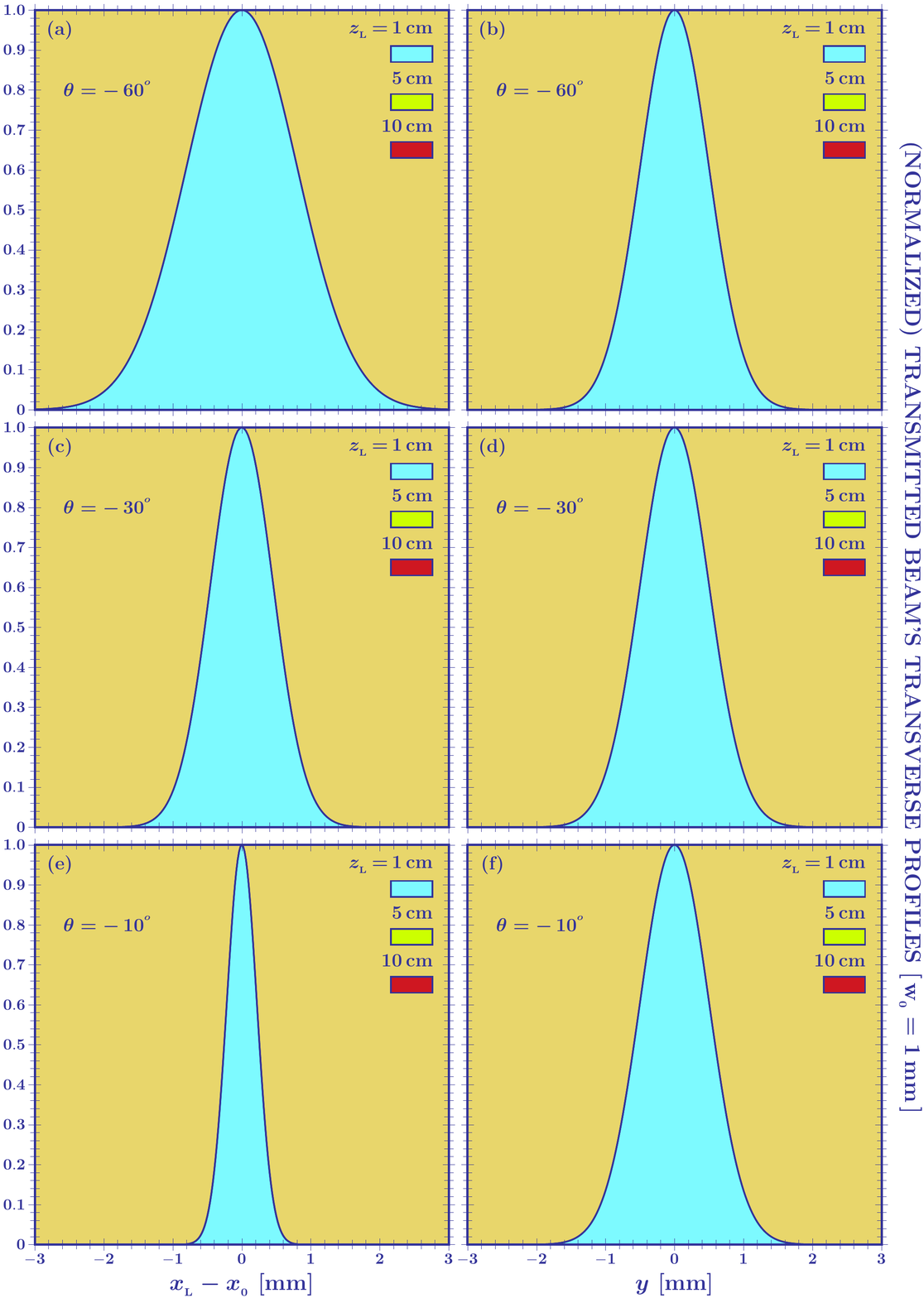}{The same of Figure 2 for a beam waist of $1$ mm.\label{fig3}}}

\def\figurefour{
\WideFigureSideCaption{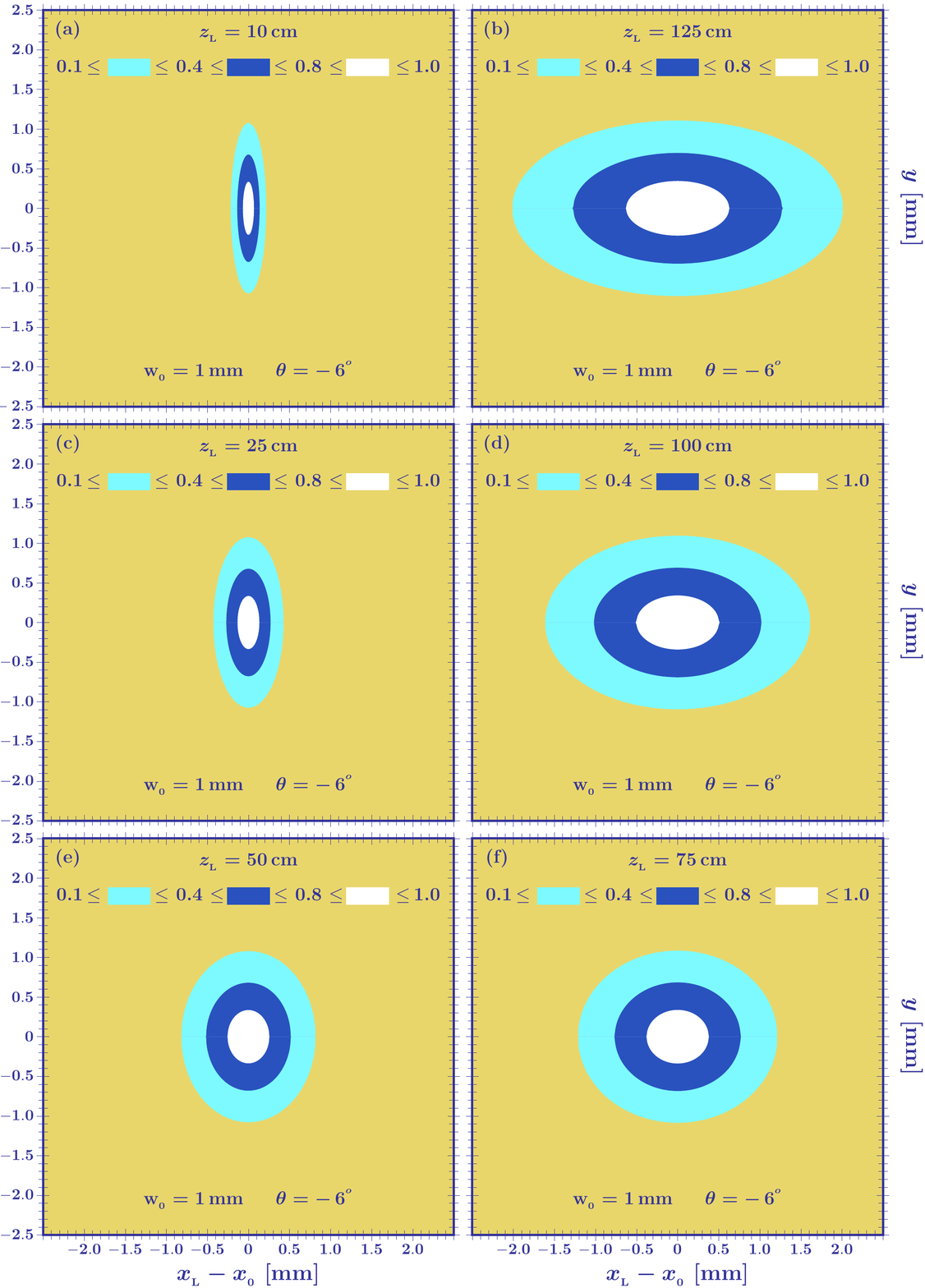}{Contour plots of the lower transmitted beam  for an incident Gaussian laser with wavelength of $633$ nm and a beam waist of $1$ mm, are plotted  for and incidence angles of $-\,6^{^{o}}$ and different axial positions. \label{fig4}}}

\def\figurefive{
\WideFigureSideCaption{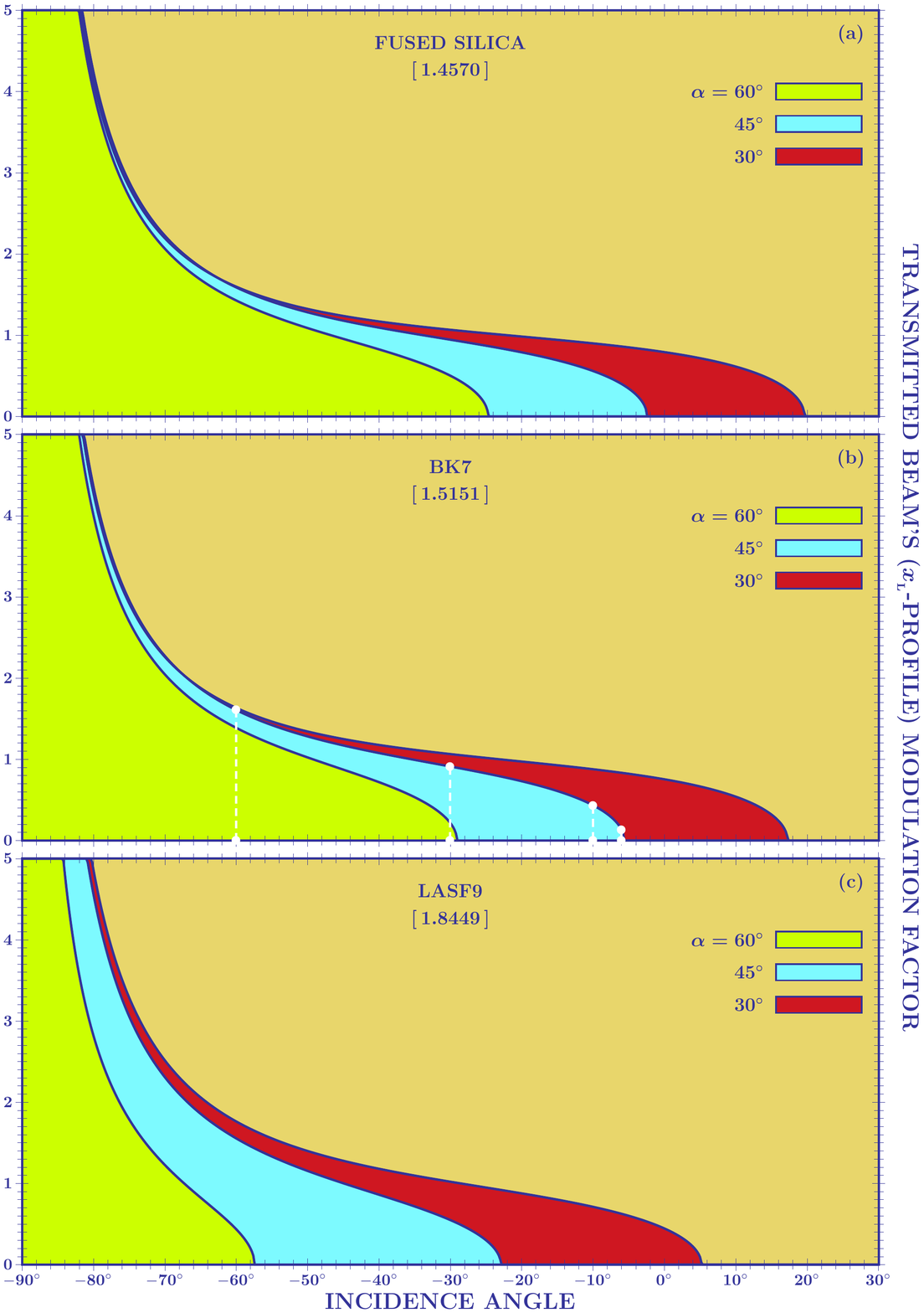}{The analytic modulation factor is plotted for different dielectrics (Fused silica, BK7, LASF9) and  geometrical $\alpha$ angles ($30^{^{o}}$,  $45^{^{o}}$,  and  $60^{^{o}}$) as a function of the incidence angle. In (b) for $\alpha=\pi/4$, the numerical data show an excellent agreement with the closed expression given for the modulation factor.  
\label{fig5}}}

\def\figuresix{
\WideFigureSideCaption{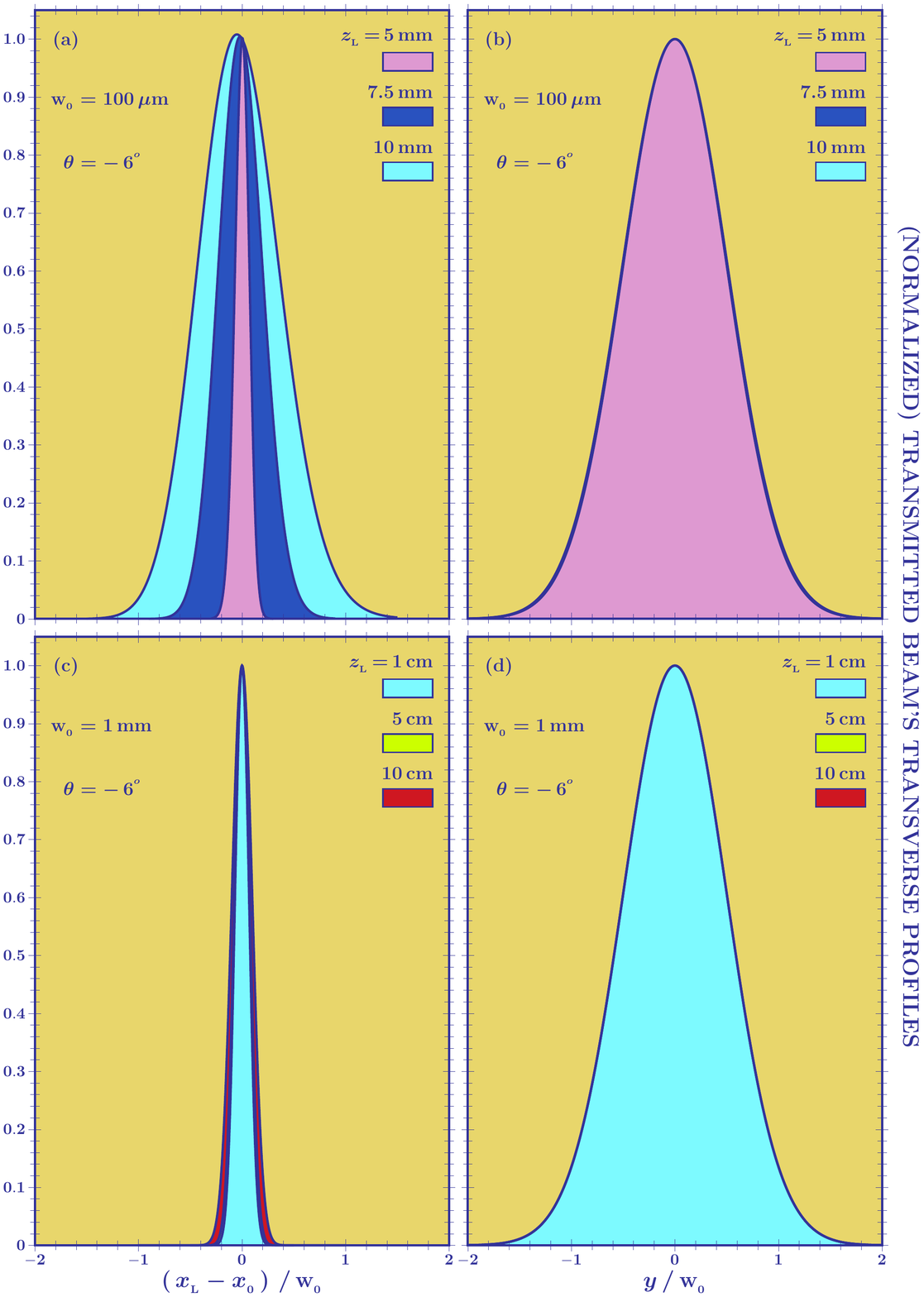}{The same of Figure 2 for an incidence angle of $-\,6^{^{o}}$.
\label{fig6}}}

\logo{
\colorbox{DarkGoldenrod}{\color{white}$\mathbf{\Sigma\hspace*{0.06cm} \delta \hspace*{0.04cm} \Lambda}$}
}

\journal{\shadowtext{\textbf{\color{DarkRed} Laser Physics Letters}} \,\,\, \textbf{17}, 116001-10 (2020).}

\titlelines{1}
\title{Laser planar trapping}

\imgbgabstract{We investigate the phenomenon of the transverse breaking of symmetry of Gaussian lasers transmitted through dielectrics,  showing for which incidence conditions and beam parameters it can be observed in optical experiments.  The numerical  analysis, done by using the integral form of the transmitted beam at the lower interface of a dielectric prism, shows, for incidence approaching the critical region, a  laser planar trapping. By using an analytic approximation for the intensity of lower transmitted beam, we model the wave front curvature  by a closed expression which depends on the refractive index, on the geometrical properties  of the dielectric block, and, finally, on the incidence angle of the incoming beam. The analytic formula shows an excellent agreement  with the numerical analysis and it could be useful in future  experimental 
implementations.}

\author{
\names{Stefano De Leo}
\affiliation{Department of Applied Mathematics, State University of Campinas, Brazil}
\email{deleo@ime.unicamp.br}
}

\begin{document}

\sdlmaketitle

\section{Introduction}

\figureone

\figuretwo

\figurethree

\figurefour

\figurefive

\figuresix

In free space, the wavelength, the position of the beam waist and the waist diameter completely describe 
the behavior of a Gaussian optical beam\cite{born,sharma,saleh}. When a beam passes through a dielectric interface, the wave front curvature is changed, resulting in new values of the waist diameter on the output side of the interface.  In this paper, we aim to analyse the incidence conditions and beam parameters which maximize 
the breaking of the transverse symmetry  of a Gaussian optical beam transmitted at the lower  (dielectric/air) interface of a dielectric prism, see Fig.\,1(a). The analysis, first done by numerically integrating the transmitted beam, will then be completed  by giving an analytic expression for the modulating factor of the  waist diameter parallel to the incidence plane.   
For the purpose of this article, it is convenient to write  the incident electric field in its integral 
form \cite{s2013,s2014},
\begin{equation}
\label{inc1}
E\inc(\mathbf{r}) = E_{\0}\,\int\hspace*{-.1cm} \d k_{_x} \,\d k_{_y} \,\,g(k_{_x},k_{_y}) \, \,
e^{{\,i \,\boldsymbol{k}\,\cdot\, \mathbf{r}}}\,\,,
 \end{equation}
where  
\[
g\left(k_{_x},k_{_y}\right) = \frac{\wo^{\2}}{4\,\pi}\, \exp \left[\,-
\,\left(\,k_{_x}^{^2} +k_{_y}^{^{2}}\,\right)\,\frac{\wo^{\2}}{4}\,\right]
\]
is the Gaussian wave number distribution, $\wo$ the laser beam waist, $|\boldsymbol{k}|=2\pi/\lambda$, and $\lambda$ the wavelength. By using the paraxial approximation\cite{saleh}, 
\[ k_z\approx |\boldsymbol{k}| - 
(\,k_{_x}^{^{2}}+k_{_y}^{^{2}}\,)\,/\,2\,|\boldsymbol{k}|\,\,,\] 
the previous integral can analytically  be solved leading to a Gaussian beam,
\begin{equation}
\label{inc2}
E\inc(\mathbf{r})= \frac{E_{\0}\,e^{{\,i\,|\boldsymbol{k}|\,z}}}{1 + i\,z/z_{_0}}\,\exp \left[\,-\,\,\frac{x^{^2}+y^{^{2}}}{\mbox{w}_{\0}^{^2}\, (1 + i\,z / z_{_{0}})}\,\right]\,\,,
\end{equation}
where $z_{_{0}}=\pi\,\wo^{^2}/\lambda$ is the Rayleigh axial range. The intensity, $I\inc =\left|E\inc\right|^{^{2}}$, 
\begin{equation}
\label{int}
I\inc  = I_{\0} \, \left[\frac{\wo}{\w(z)}\right]^{^2} \, \exp \left[\,- \,2 \,\,\frac{ x^{^2} + y^{^2}}{\w^{^2}(z)}\,\right]\,\,,
 \end{equation}
 where $I_{\0}=|E_{\0}|^{^{2}}$ and 
 \[ \w(z) = \wo \, \sqrt{1 + (z/z_{_{0}})^{^2}}\,\,,\]
  shows the well-known transverse symmetry of Gaussian beams.

This paper was prepared with the aim to show when this transverse symmetry is broken, to understand  the reason of such a breaking of symmetry and, once numerically confirmed, to give an analytic closed expression to be used in future experimental implementations. As we shall see in detail later, this breaking of symmetry  acts on the transverse component positioned in the plane of incidence, and, for critical incidence, generates a planar trapping of the beam in the plane perpendicular to the incidence one. It is important to observe  that the phenomenon presented in this article can also be seen in a reversed order. Indeed, by an appropriate choice of the geometry and refractive index of the dielectric block and of the incidence angle,  this effect can also be used  to correct transverse asymmetries in optical beams.

\section{The lower transmitted beam}

When a plane electromagnetic
wave encounters a plane interface between  two transparent media, reflected and transmitted waves are generated at such an interface. The angles which these waves form with the normal to such an interface are determined
by the reflection and Snell laws and the amplitudes described by the Fresnel coefficients\cite{born,sharma,saleh}. For Transverse Electric (TE) and Transverse Magnetic (TM) waves, the Fresnel coefficients at the left (air/dielectric) interface, $\xt$-$\zt$ system in Fig.\,1(b), can be calculated by solving the matrix continuity equation \cite{s2015}
\begin{equation}
\label{Fre1a}
\left(
\begin{array}{cc}
1 & 1\\
k_{_{\zt}} & -\, k_{_{\zt}}\end{array}
\right)
\left[\begin{array}{c} 1 \\
\widetilde{R}^{^{[\sigma]}}\end{array}\right]
 =  \left[\begin{array}{c} a_{_\sigma} \\ q_{_{\zt}}/a_{_\sigma}
\end{array}\right] \,\,\widetilde{T}^{^{[\sigma]}}\,\,,
\end{equation}
where $\left\{\,a_{_\mathrm{TE}}\,,\,a_{_\mathrm{TM}}\right\}=\left\{\,1\,,\,n\,\right\}\,$, 
$k_{_{\zt}}= -\,k_x\,\sin\theta +k_z \,\cos\theta$, and $q_{_{\zt}}=\sqrt{(n^{\2}-1) |\boldsymbol{k}|^{^{2}}+ k_{_{\zt}}^{^{2}}}$.
The Fresnel coefficients depend on the wave number components appearing in the Gaussian distribution, 
\begin{equation}
\label{Fre1b}
\widetilde{T}^{^{[\sigma]}}(k_{_x},k_{_y}) \, =\,
2\,k_{_{\zt}}\,/\,\left(\,a_{_\sigma}\,k_{_{\zt}}\,+\,q_{_{\zt}}/\,a_{_\sigma}\,\right)\,\,,
\end{equation} 
and reproduce the well known angular Fresnel coefficients when calculated at the center of the wave number  distribution \cite{born,sharma,saleh}, i.e.
\begin{equation}
\label{Fre1c}
\widetilde{T}^{^{[\sigma]}}(0,0) \, =\,
2\,\cos\theta\,/\,\left(\,a_{_\sigma}\,\cos\theta\,+\,n\,\cos\psi/\,a_{_\sigma}\,\right)\,\,.
\end{equation} 
At the lower dielectric/air interface,  $\xs$-$\zs$ system in Fig.\,1(b), the Fresnel continuity equations
 become
\begin{eqnarray}
\label{Fre2a}
\left(
\begin{array}{cc}
a_{_\sigma} & a_{_\sigma}\\ & \\
\displaystyle{\frac{q_{_{\zs}}}{a_{_\sigma}}} & -\,\displaystyle{\frac{q_{_{\zs}}}{a_{_\sigma}}}\end{array}
\right)
\left[\begin{array}{c} e^{\,i\,q_{_{\zs}}h} \\ \\
R_{_{*}}^{^{[\sigma]}} e^{\,-\,i\,q_{_{\zs}}h}\end{array}\right]\, 
 & = & \nonumber \\
   \left[\begin{array}{c} a_{_\sigma} \\ \\ \displaystyle{\frac{q_{_{\zs}}}{a_{_\sigma}}}
\end{array}\right] \,\,T_{_{*}}^{^{[\sigma]}}e^{\,i\,k_{_{\zs}}h} & & 
\end{eqnarray} 
where $q_{_{\zs}}= -\,q_{\xt}\, \sin\alpha +q_{\zt}\, \cos\alpha$ (note that  $q_{\xt}=k_{\xt}$), 
$k_{_{\zs}}= \sqrt{(1-n^{\2}) |\boldsymbol{k}|^{^{2}}+ q_{_{\zs}}^{^{2}}}$, and $h$ is the distance along the $\zs$ axis between the lower (dielectric/air) discontinuity and the origin of the axes fixed, in our analysis,  at the point in which the incident beam touches the left 
(air/dielectric) interface, see Fig.\,1(a).  

This choice implies for an experimental implementation to set the  position of the beam waist at the point in which the incident beam crosses the left (air/dielectric) interface. 
Solving Eqs.\,(\ref{Fre2a}), we find
\begin{equation}
\label{Fre2b}
T_{_{*}}^{^{[\sigma]}}(k_{_x},k_{_y}) =
\frac{2\,q_{_{\zs}}\,e^{\,i\,(q_{_{\zs}}-k_{_{\zs}})\,h}}{a_{_\sigma}\,k_{_{\zs}}\,+\,q_{_{\zs}}/\,a_{_\sigma}}\,\,,
\end{equation} 
which, at the center of the Gaussian distribution, become
\begin{equation}
\label{Fre2c}
T_{_{*}}^{^{[\sigma]}}(0,0) =
\frac{2\,n\,\cos\varphi\,e^{\,i\,(n\,\cos\varphi-\cos\phi)\,|\boldsymbol{k}|\,h}}{a_{_\sigma}\,\cos\phi\,
+\,n\,\cos\varphi/\,a_{_\sigma}}\,\,.
\end{equation} 
Finally, the integral form of the lower transmitted beam is given by 
\begin{eqnarray}
\label{LT}
E\L^{^{[\sigma]}}(\xs,y,\zs) &=& E_{\0}\,\int\hspace*{-.1cm} \d k_{_x} \,\d k_{_y} \,\,
T\L^{^{[\sigma]}}\left(k_{_x},k_{_y}\right)\, g\left(k_{_x},k_{_y}\right)\,\times \nonumber \\  
 & & \exp[\,i\,(\,k_{_{\xs}}\,\xs+k_{_y}\,y+k_{_{\zs}}\,\zs\,]\,\,,
 \end{eqnarray}
where $T\L^{^{[\sigma]}}= \widetilde{T}^{^{[\sigma]}}\,T_{_{*}}^{^{[\sigma]}}$. 

In order to check the transverse symmetry of the lower transmitted beam, it is convenient to rewrite the optical spatial phase in terms of its proper coordinates $x\L$-$\,z\L$ system, drawn in Fig.\,1(b),
\[ (k_{_{\xs}}\cos\phi\,-\,k_{_{\zs}}\sin\phi)x\L+k_{_y}\,y+
(k_{_{\xs}}\sin\phi+k_{_{\zs}}\cos\phi)z\L\,\,,
\]
and then numerically calculate the intensity.

The $x\L/y$ transverse profiles of the normalized lower transmitted beam intensity 
\[\left| E\L^{^{[\sigma]}}(x\L,y,z\L)\,/\, E\L^{^{[\sigma]}}(x_{\0},0,z\L)\right|^{^{2}}\,\,,   \]
 where $x_{\0}=h\,\sin(\varphi-\phi)/\cos\varphi\,$ is the center of the beam predicted by the Snell law
 ($\sin\theta=n\,\sin\psi$, $\varphi=\psi+\alpha$,  $n\,\sin\varphi=\sin\phi$), are plotted in Fig.\,2  for a  laser, with wave length of $633\,\mathrm{nm}$ and  $\wo=100\,\mu\mathrm{m}$, whose axial direction forms incidence angles of $-\,60^{^{o}}$, $-\,30^{^{o}}$, and   $-\,10^{^{o}}$  with the normal to the left (air/dielectric) interface of a BK7 ($n=1.5151$) dielectric prism with $\alpha=\pi/4$. For each incidence angle,  the intensity of the beam is plotted for three axial values, i.e. $z\L=1,\,5\,,10\,\mathrm{cm}$.   The transverse profiles appearing in Fig.\,3 refer to a laser with $\wo=1\,\mathrm{mm}$.

The numerical analysis shows that the $y$-profiles of the lower transmitted beam are practically unchanged with respect to the incident one. So, the first conclusion is  that,  for short propagation in the dielectric,  the wave front of the beam does not change in the direction perpendicular to the plane of incidence.  For what concerns the $x\L$-profiles,  the data plotted in Figs.\,2 and 3 clearly manifest  a breaking of symmetry in the transverse component parallel to the plane of incidence.

To quantify this breaking of symmetry, we first use the numerical data of the profiles plotted in Fig.\,3.
In this case, due to the fact that the axial dependence does not play a significant role,  by taking, from the numerical data, the ratio between the $x\L$ and the $y$ profile, we immediately obtain   
\begin{equation}
\left[\begin{array}{c} \widetilde{\w}_{\0,\m\6\0^{\o}}\\\widetilde{\w}_{\0,\m\3\0^{\o}}\\
\widetilde{\w}_{\0,\m\1\0^{\o}}\end{array}\right] = \left[\begin{array}{c} 1.60\\0.91\\0.43 
\end{array}\right]\,\mathrm{mm}\,\,.
\end{equation}
The modulation factors $1.60$, $0.91$, and $0.43$  are in excellent agreement with the axial behaviour showed
for the $x\L$-profiles in Fig.\,2, where $\wo=100\,\mu \mathrm{m}$. This clearly suggests that the breaking of symmetry in the transverse profiles parallel to the incidence plane does not depend on the choice of $\wo$. One point we should stress is that the planar trapping in the plane perpendicular to the incidence plane is found for incidence 
angles  approaching  the critical angle, $\varphi_{_{\mathrm{cri}}}=\arcsin(1/n)$  
which implies  $\theta_{_{\mathrm{cri}}}=-\,5.6^{^{o}}$, and, due to the increase of the beam divergence, it is detectable  for axial position of the camera  near to the lower interface. The contour plots shown in Fig.\,4 
refer to  an incident Gaussian beam with $\wo = 1\,\mathrm{mm}$ and incidence angle $\theta=-\,6^{^{o}}$. The plots  manifest a  planar trapping at an axial distance $z\L \leq 10\,\mathrm{cm}$. The axial behaviour of $x\L$ profile of  the lower transmitted beam  is compatible with 
\begin{equation}
\widetilde{\w}_{\0,\m\6^{\o}}\,=\,0.13\,\mathrm{mm}\,\,.
\end{equation}

\section{The modulation factor}

In order to obtain an analytic expression for the beam divergence of the lower transmitted beam  in the plane parallel to the incidence plane, we have to calculate the intensity of the lower transmitted beam by approximating  the transmission coefficient, which appears in the modulated wave number distribution, by 
$T\L^{^{[\sigma]}}(0,0)$  and by using the Taylor expansion  of the optical phase up to the second order in $k_x$ and $k_y$.

The spatial phase of the lower transmitted beam can then be approximated by 
\begin{equation*}
|\boldsymbol{k}|\,z\L\,+\,k_{_{x}}\,\frac{x\L-x_{\0}}{\rho\L}\,\,-\,\,\frac{k^{^{2}}_{_{x}}}{2\,|\boldsymbol{k}|}\,\,
\frac{z\L}{\rho\L^{^{2}}}\,\,+\,\,
k_{_{y}}\,y\,\,-\,\,\frac{k^{^{2}}_{_{y}}}{2\,|\boldsymbol{k}|}\,\,z\L\,\,,
\end{equation*}
where $x_{\0}=h\,\sin(\varphi-\phi)/\cos\varphi$ is the geometrical shift predicted by the Snell law and 
\begin{equation}
\label{mfl}
\rho\L = \frac{\cos\psi\,\cos\phi}{\cos\theta\,\cos\varphi}
\end{equation}
is an angular ratio which is an implicit function of the incidence angle $\theta$ and of the refraction index $n$ (which also contains an implicit dependence on $\lambda$).    From the approximated spatial phase, we immediately observe the breaking of symmetry between the the transverse components of the lower transmitted beam.

The approximations done for transmission coefficient and for the optical phase allow us to analytic integrate Eq.\,(\ref{LT}) and to obtain a closed  expression for the lower transmitted beam and consequently for its intensity,
\begin{eqnarray}
\label{IL1}
I\L^{^{[\sigma]}}& = & I_{\0} \,\,\rho\L\,\,\left|\,T\L^{^{[\sigma]}}(0,0)\,\right|^{^{2}}\,\frac{\wo^{^2}}{\w(z\L)\,
\w(z\L/\rho\L^{^2})} \,\times \\
& &
\exp \left[- \,2\,\frac{(\,x\L\,-\,x_{\0}\,)^{^2}/\rho\L^{^{2}}\,)}{\w^{^2}(z\L/\rho\L^{^2})}\,\right]
 \exp \left[- \,2\,\frac{ y^{^{2}} }{\w^{^2}(z\L)}\right]\,\,.\nonumber
 \end{eqnarray}
Introducing
\begin{eqnarray}
\widetilde{\w}(z\L)&=&\rho\L\,\w(z\L/\rho\L^{^2})\nonumber\\
&=&\rho\L\,\wo\,\sqrt{1 + (z\L/\rho\L^{^{2}}z_{_{0}})^{^2}}
\nonumber \\
 &=&   \widetilde{\w}_{\0} \, \sqrt{1 + (z\,/\,\widetilde{z}_{_{0}})^{^2}}\,\,,
\end{eqnarray}
where $\,\widetilde{z}_{\0}=\pi\,\widetilde{\w}_{\0}^{^2}/\lambda$, we can rewrite the intensity in terms of the proper transverse waists of the beam, i.e. $\w_{\0}$ for the transverse component perpendicular to the plane  of incidence and  $\widetilde{\w}_{\0}$ for the transverse one parallel to the plane of incidence,
\begin{eqnarray}
\label{IL2}
I\L^{^{[\sigma]}}& = & I_{\0} \,\,\rho\L\,\,\left|\,T\L^{^{[\sigma]}}(0,0)\,\right|^{^{2}}\,\frac{\wo\widetilde{\w}_{_{0}}}{\w(z\L)\,
\widetilde{\w}(z\L)} \,\times \\
 & &   \exp \left[- \,2\,
\frac{(\,x\L\,-\,x_{\0}\,)^{^2}}{\widetilde{\w}^{^2}(z\L)}\,\right]\,\exp \left\{- \,2\,\frac{y^{^{2}} }{\w^{^2}(z\L)}\right]\,\,. \nonumber 
 \end{eqnarray}
 The modulation factor $\rho\L$ is plotted in Fig.\,5 and shows an excellent agreement with the numerical data obtained in Section II.

Once obtained the analytical expression for the modulation factor $\rho\L$, it is also possible to calculate the incidence angle for which we recover the transverse symmetry between. Indeed, by solving the equation 
\[\rho\L=1\,\,,\] which implies
\[\cos^{\2}\psi\sym[1\,-\,n^{\2}\,\sin^{\2}\varphi\sym]=(1\,-\,n^{\2}\,\sin^{\2}\psi\sym)\,\cos^{\2}\varphi\sym\,\,,
\]
we immediately get $ \cos^{\2}\psi\sym\,\,=\,\cos^{\2}\varphi\sym$  and consequently 
$\psi\sym\,=\,-\,\alpha\,/\,2$. This means that the transverse symmetry  is recovered for the following incidence angle 
\begin{equation} 
\theta\sym=\,-\,\arcsin\left(\,n\,\sin\frac{\alpha}{2}\,\right)\,\,.
\end{equation}
For  Fused Silica, BK7, and LASF9 prisms, the incidence angles for which we recover the transverse symmetry, in the case of an incoming beam with a wavelength    of 633 nm, are 
\begin{equation*}
\left( \begin{array}{c}
\theta^{^{\mathrm{[FuSi]}}}\sym\\
\theta^{^{\mathrm{[BK7]}}}\sym\\
\theta^{^{\mathrm{[LAS]}}}\sym\end{array}\right) =
-\,\left(\begin{array}{r}
22.15^{^{o}}_{_{\pi/6}}\,,\,33.89^{^{o}}_{_{\pi/4}}\,,\,46.76^{^{o}}_{_{\pi/3}}\\
23.09^{^{o}}_{_{\pi/6}}\,,\,35.44^{^{o}}_{_{\pi/4}}\,,\,49.25^{^{o}}_{_{\pi/3}}\\
28.52^{^{o}}_{_{\pi/6}}\,,\,44.91^{^{o}}_{_{\pi/4}}\,,\,67.29^{^{o}}_{_{\pi/3}}\end{array}\right)\,\,,
\end{equation*}
where the lower script angles indicate the $\alpha$ angle of the prism. 

The modulation factor $\rho\L$ goes to zero when the incidence angle approaches the critical angle, 
\[
\theta\cri\,=\,\arcsin \left[n\,\sin\left( \arcsin\frac{1}{n}\,-\,\alpha\right)\right]\,\,.
\]
For  Fused Silica, BK7, and LASF9 prisms, the critical angle are
\begin{equation*}
\left( \begin{array}{c}
\theta^{^{\mathrm{[FuSi]}}}\cri\\
\theta^{^{\mathrm{[BK7]}}}\cri\\
\theta^{^{\mathrm{[LAS]}}}\cri\end{array}\right) =
\left(\begin{array}{r}
19.65^{^{o}}_{_{\pi/6}}\,,\,-\,2.42^{^{o}}_{_{\pi/4}}\,,\,-\,24.69^{^{o}}_{_{\pi/3}}\\
17.27^{^{o}}_{_{\pi/6}}\,,\,-\,5.61^{^{o}}_{_{\pi/4}}\,,\,-\,29.06^{^{o}}_{_{\pi/3}}\\
5.21^{^{o}}_{_{\pi/6}}\,,\,-\,22.90^{^{o}}_{_{\pi/4}}\,,\,-\,57.42^{^{o}}_{_{\pi/3}}\end{array}\right)\,\,.
\end{equation*}
Incidence near the critical angle generates the phenomenon of planar trapping. For a beam waist of 1 mm, the planar trapping is evident in the contour plots of Fig.\,4(a). In Fig.\,6, it is shown what happens when  we reduce the beam waist up to 100 $\mu$m. The center  of the transverse $x\L$ profiles moves when we  increase the axial distance $z\L$ and this is a clear evidence of angular deviations\cite{Ang1,Ang2,Ang3,Ang4}.

\section{Conclusions}

The main objective of the study presented in this paper was to obtain an analytical formula for the waist diameter modification of an optical Gaussian beam transmitted through dielectrics. The analysis was done for the transmission from an air/dielectric and a dielectric/air interface forming an angle $\alpha$. For experimental implementations,  the prisms found in commerce could be used to test our prediction for $\alpha=\pi/4$. 
 
The analytical formula for the modulation factor  of the waist diameter
\begin{equation}
\label{modf}
\widetilde{\w}_{\0}\,=\,\frac{\cos \psi\,\cos\phi}{\cos \theta \cos\varphi}
\,\wo\,\,,
\end{equation}
valid for $\theta\leq\theta\cri$, was found by using the Taylor expansion of the optical phase up to its second order and by using the Fresnel coefficients calculated at the center of the Guassian distribution.  It depends on the geometrical properties of the dielectric, parameter $\alpha$, on the incidence angle, parameter $\theta$, and on the refractive index, parameter $n$, which also depends on the wavelength $\lambda$ of the beam (Sellmeier equations) and shows
an excellent agreement, see Fig.\,5, with the numerical data extracted by the plots of Figs.\, 2,3, and 4.

Once obtained the closed expression for the modulation factor, Eq.\,(\ref{modf}), it can be used in optical experiments to modulate Gaussian lasers or correct eventual transverse asymmetries of the beam. For example,
by using a BK7 prism (with $\alpha=\pi/4$) depending of the wavelength of the incident laser we can easily obtain a reduction to one  half or an amplification of a factor two by using the following incidence angle,
\begin{equation*}
\left(\begin{array}{c}
 \theta^{^{\mathrm{[532\,nm]}}}_{_{0.5}}\,\,\,  \theta^{^{\mathrm{[532\,nm]}}}_{_{2.0}}\\ \\  
\theta^{^{\mathrm{[633\,nm]}}}_{_{0.5}}\,\,\,  \theta^{^{\mathrm{[633\,nm]}}}_{_{2.0}}  
\end{array}\right)_{_{\mathrm{[BK7]}}} = -\,
\left(\begin{array}{c}
11.85^{^{o}}_{_{\pi/4}}\,\,\,66.85^{^{o}}_{_{\pi/4}}\\ \\
11.60^{^{o}}_{_{\pi/4}}\,\,\,66.88^{^{o}}_{_{\pi/4}}\end{array}\right)\,\,.
\end{equation*}
To obtain an optical beam with the same modulation factor in the transverse components, we can use a second prism rotated by an angle of $\pm\,\pi/2$ along the $z$-axis, see FIg.\,1. The lower transmitted beam is then forced to pass trough the second prism with the same incidence angle used in the first prism. In this way, we compensate the asymmetry created by the first transmission. Finally, we observe that Eq.\,(\ref{modf}) is also useful to prove that there is no modulation in the upper transmitted beam. Eq.\,(\ref{modf}) contains the product 
of the ratio between the transmitted and incident wave number component along the normal to the air/dielectric and dielectric/air interfaces. For transmission trough the left (air/dielectric) and lower (dielectric/air) interfaces, we have found $(\cos\psi/\cos\theta)\,\times\,(\cos\phi/\cos\varphi)$. Consequently, for transmission through the left (air/dielectric) and upper (dielectric/air) interface, we expect   $(\cos\psi/\cos\theta)\,\times\, (\cos\theta/\cos\psi)\,=\,1$. We hope that the theoretical prediction of the waist modulation reported in this paper could find a confirmation in future experimental implementations.

\subsection*{Acknowledgements}

The author (S.D.L) thanks the CNPq (grant 2018/303911) and Fapesp (grant 2019/06382-9) for financial support.

\end{document}